%% file: preprint2027.tex
\documentclass[letterpaper]{article} 
\usepackage{aaai2027}  
\usepackage[hyphens]{url}  
\usepackage{graphicx} 
\usepackage{natbib}  
\usepackage{caption} 
\usepackage{algorithm}
\usepackage{algpseudocode}

\usepackage{microtype}
\usepackage{booktabs}
\usepackage{amsmath, amssymb, amsfonts, mathtools, bm}
\usepackage{amsthm}
\usepackage{booktabs}   
\usepackage{makecell}   
\usepackage[table]{xcolor} 
\usepackage{threeparttable}
\usepackage{booktabs, multirow, tabularx}
\usepackage{array}
\usepackage[most]{tcolorbox}
\tcbuselibrary{breakable}
\usepackage{graphicx, subcaption}

\usepackage{cleveref}

\usepackage{xcolor, xspace, enumitem, microtype}
\newcolumntype{Y}{>{\centering\arraybackslash}X}
\usepackage{dsfont}

\usepackage{lineno}

\definecolor{darkblue}{rgb}{0, 0, 0.5}
\usepackage[most]{tcolorbox}
\usepackage{listings}

\lstdefinestyle{promptlisting}{
  basicstyle=\ttfamily\footnotesize,
  columns=fullflexible,
  keepspaces=true,
  showstringspaces=false,
  breaklines=true,
  breakatwhitespace=false,
  tabsize=2,
  upquote=true,
  literate={→}{{$\rightarrow$}}1 {—}{{--}}1
}

\newtcblisting{promptbox}[1]{
  enhanced,
  breakable,
  listing only,
  listing options={style=promptlisting},
  colback=black!3!white,
  colframe=black!70!white,
  boxrule=0.5pt,
  arc=1mm,
  boxsep=1pt,
  left=1mm,
  right=1mm,
  top=1mm,
  bottom=1mm,
  title={#1},
  fonttitle=\bfseries\small,
  before skip=6pt,
  after skip=8pt
}

\usepackage{booktabs}

\usepackage{newfloat}
\usepackage{listings}
\DeclareCaptionStyle{ruled}{labelfont=normalfont,labelsep=colon,strut=off} 
\floatstyle{ruled}
\newfloat{listing}{tb}{lst}{}
\floatname{listing}{Listing}

\usepackage{booktabs}

\newcommand{\trajt}[1]{\tau^i_{#1}}

\newcommand{\policyego}[1]{\pi^i_{#1}}

\newcommand{\policyteam}[2]{\pi^{#2}_{#1}}

\newcommand{\stateinitdist}{\rho_{0}^{s}}

\title{Bayesian Partner Modelling enables Adaptive Replanning for LLM Coordination}
\author{
    Harsh Goel\textsuperscript{\rm 1}\equalcontrib, Aditya Sai Ellendula\textsuperscript{\rm 1}\equalcontrib, Vaishnav Tadiparthi\textsuperscript{\rm 2}, Ehsan Moradi Pari\textsuperscript{\rm 2}, Hossein Nourkhiz Mahjoub\textsuperscript{\rm 2}, Sandeep P. Chinchali\textsuperscript{\rm 1}
}
\affiliations{

    \textsuperscript{\rm 1}The University of Texas at Austin
    \textsuperscript{\rm 2} Honda Research Institute, USA\\
}
\nocopyright

\begin{document}

\maketitle

\begin{abstract}
Multi-agent Large Language Model (LLM) systems often struggle to collaborate with new teammates whose strategies shift mid-task. Because agents execute multi-step or temporally extended skills, they frequently continue executing outdated plans long after public evidence shows that a partner has changed its skill. Existing methods either treat partner tracking as passive context—leaving the agent aware of the shift but slow to act—or replan indiscriminately. We introduce BayesBeliefAgent, which pairs a hierarchical LLM planner with a Bayesian tracking module. Rather than replanning constantly, our agent interrupts its current skill only when a partner’s actions directly contradict the inferred skill. Beyond standard reward, we evaluate performance using replanning efficiency and the belief–action gap: the fraction of total decisions where an agent with a correct partner estimate executes a non-complementary skill. Across benchmark Overcooked environments, contradiction-conditioned control drastically narrows this belief–action gap while requiring an order of magnitude fewer replans than heuristic methods.
\end{abstract}

\input{tex/intro}
\input{tex/related_works}

\input{tex/method}
\input{tex/experiments}

\section{Future Work and Conclusion}

\textbf{Limitations.}
A key limitation is that \textsc{BayesBeliefAgent} reasons over a fixed set of coarse, manually specified skills. A natural next step is to learn this skill space and represent finer-grained structure, including subgoal dependencies and open-ended skill discovery. Posterior-gated interruption is also less effective in tightly coupled environments such as Forced Coordination. Improvements in these settings will likely require stronger LLM-based planners, as well as methods that jointly adapt belief estimation and action selection. Additionally, evaluations across a larger number of partners and human partners whose strategies change within an episode remain an important direction for future work.

\textbf{Conclusion.} In this work, we introduce \textsc{BayesBeliefAgent}, a hierarchical LLM agent that maintains an online Bayesian belief over a teammate’s skills and uses unexpected observations to decide when an ongoing skill should be interrupted and revised. This design directly addresses the belief–action gap: an agent may correctly infer what its teammate is doing while continuing to execute a stale or duplicative skill. By connecting partner belief to selective replanning, \textsc{BayesBeliefAgent} reduces duplicate-role behavior, improves recovery after contradictory observations, and achieves rewards comparable to generic replanning with substantially fewer replans. Ablations further show that the posterior is most useful as a control signal rather than merely as additional planner context. These findings suggest that collaborative theory of mind should be leveraged for timely changes in behavior.

\newpage
\bibliography{aaai2027}
\newpage
\input{tex/appendix}

\end{document}

%% file: tex/intro.tex
\section{Introduction}

Imagine a student who realizes that their professor has changed the direction of a joint project, yet continues carrying out an analysis that is no longer useful instead of stopping and replanning new experiments. Large language model (LLM) agents face an analogous challenge when collaborating in sequential decision-making tasks: they must coordinate with teammates, whose behavior may change during execution~\citep{zhang2024proagent, agashe2025llmcoord}. In such settings, coordination failures do not always arise because an agent misunderstands its partner’s intent: a capability commonly associated with Theory of Mind~\citep{premack1978theory, byomtom2013, shu2021agent, gao2024vision}. Rather, they arise because the agent fails to replan at the right time.

\begin{figure}[tbp]
    \centering
    \includegraphics[width=0.49\textwidth]{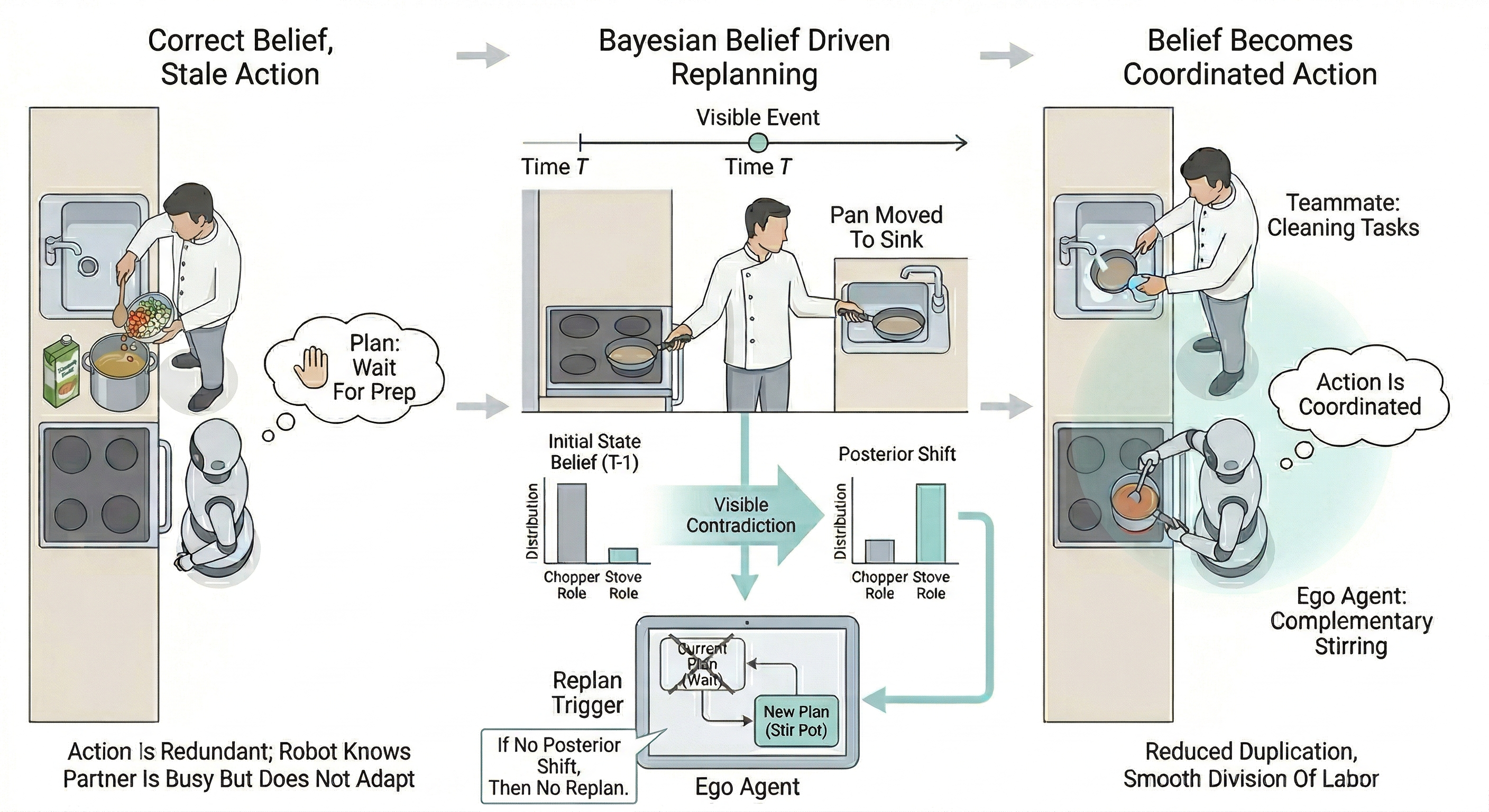}
    \caption{
    \textbf{Belief alone is not enough.}
    An LLM agent may correctly model its teammate yet still act redundantly if it does not revise its plan when partner behavior changes.
    \textsc{BayesBeliefAgent} addresses this by maintaining a Bayesian posterior over latent teammate coordination skills, detecting visible contradictions, and triggering replanning when the current plan becomes stale.
    This turns teammate belief into coordinated, complementary action.
    }
    \label{fig:teaser}
    \vspace{-2em}
\end{figure}

LLM agents in collaborative settings often continue executing an obsolete plan even after public evidence indicates that a teammate’s strategy has shifted. Recent LLM-based coordination methods address this problem by incorporating inferred partner roles into the planning context at inference time~\citep{zhang2024proagent, collaboverovercooked, recollab}. However, accurate partner inference does not necessarily translate into effective coordination. As observed by \citet{riemer2025tom}, LLM agents can exhibit a literal--functional Theory of Mind gap: they may correctly infer a partner’s role or goal yet fail to adapt their own behavior accordingly. In sequential tasks, this failure can arise when an agent updates its belief about a teammate’s new role but continues executing a skill selected under its previous belief. Its partner model has changed, but its control policy remains committed to an obsolete course of action. This exposes a \emph{belief-to-control} gap: understanding what a partner is doing does not by itself provide a principled mechanism for deciding whether and when to interrupt ongoing execution. The central challenge is therefore not only to infer a teammate’s intent, but also to determine when new evidence is sufficiently consequential to trigger replanning. Effective coordination thus requires \emph{selective replanning}: interrupting an ongoing skill when public observations contradict the assumptions under which it was selected, while avoiding the excessive churn caused by reacting to every behavioral change. Our experiments further show that generic interruption rules are insufficient to make this decision reliably.

We address this gap with \textsc{BayesBeliefAgent}, which tracks the skill the partner is likely executing and uses that estimate to decide when to replan. For each possible partner skill, the agent predicts the low-level actions the partner would take and compares those predictions with the observed public behavior. This comparison updates its estimate of the partner’s skill. When the partner takes actions that conflict with the current estimate, the agent interrupts its ongoing skill and replans.

We evaluate BayesBeliefAgent in Overcooked and Burrito with previously unseen partners exhibiting diverse behavioral preferences \cite{wang2024zsc}. Because episodic reward is noisy and can obscure coordination failures, we additionally report four process-level metrics: the \emph{belief-action gap rate}, defined as the fraction of decision points at which the agent has a correct teammate-skill estimate but nevertheless selects or continues a non-complementary skill; \emph{contradiction-conditioned recovery}, which measures how quickly the agent selects a complementary skill after contradictory evidence; the \emph{duplicate-skill rate}, which measures how often both agents pursue the same skill; and \emph{replan efficiency}, which captures the amount of replanning required to coordinate effectively.

We compare against baselines that replan (i) periodically, (ii) after specific public events, or (iii) whenever an LLM judge recommends it, as well as a hypothesis-based partner-modeling method~\citep{hypotheticalminds}. Across Open and Ring layouts, \textsc{BayesBeliefAgent} reduces the gap between what the agent believes and how it acts. Simple triggers can achieve similar reward, but only by replanning more than 100 times per episode: roughly two orders of magnitude more often than \textsc{BayesBeliefAgent}.


Our key contributions are as follows:
\begin{itemize}
\item \textbf{Method.} We introduce \textsc{BayesBeliefAgent}, a hierarchical LLM coordination agent that uses likelihood-based contradictions under the current teammate-skill estimate to selectively interrupt and replan.
\item \textbf{Diagnostics.} We propose a diagnostic evaluation protocol that measures the belief--action gap, contradiction-conditioned recovery, and replanning efficiency, revealing coordination failures that aggregate reward obscures.
\item \textbf{Evidence.} Across Open and Ring layouts, we show that posterior-gated interruption turns partner beliefs from passive planning context into an active control signal. It reduces the belief--action gap and improves recovery after teammate strategy shifts while requiring substantially fewer replans than periodic, event-triggered, and LLM-judge baselines. We also identify the settings in which selective replanning provides limited benefit.
\end{itemize}

%% file: tex/related_works.tex
\section{Related Work}

\paragraph{Ad hoc teamwork and zero-shot coordination.}
Multi-agent Reinforcement Learning (MARL) methods for coordination have traditionally emphasized learning specialized roles \citep{liu2023contrastive, li2021celebrating, wang2020roma, goel2025r3dm}. 
The problem of coordinating with unseen partners without prior joint training is the classical ad hoc teamwork setting~\cite{stone2010adhoc}, and Overcooked-AI~\cite{carroll2019overcooked} has become its standard testbed for human-AI coordination. Prior work improves performance through \emph{training-time} partner diversity: arbitrary conventions~\cite{hu2020otherplay}, population-based play~\cite{strouse2021fcp,zhao2023mep}, and open-ended partner generation~\cite{li2023cole,wang2025rotate}. These methods require environment-specific training~\cite{overcookedv2,ogc}. Our work is complementary and inference-time: given a fixed LLM agent and an unseen partner, we ask \emph{when} an updated partner belief should interrupt the agent's plan. 

\paragraph{LLM agents for multi-agent coordination.}
LLM agents have increasingly been used in cooperative tasks because they can interpret shared goals, reason about teammate behavior, and generate high-level plans. Related works~\citep{zhang2024proagent,liu2024hla,hypotheticalminds} uses teammate-intent reasoning and hierarchical planning to support proactive cooperation in Overcooked. Recent methods incorporate partner models into decision-making, either as retrieved context~\cite{recollab} or as constraints on actions~\cite{causalplan}.  However, broader evaluations across cooperative environments show that strong task understanding does not necessarily translate into robust coordination~\cite{agashe2025llmcoord,mandi2024roco}. In particular, recent studies find that LLM agents often struggle to sustain collaboration and adapt as teammate behavior changes~\cite{collaboverovercooked,riemer2025tom}. In contrast, our work studies how updated partner beliefs should control \emph{when} an agent interrupts an ongoing skill and replans.

\paragraph{Theory of mind and partner belief modeling.}
Modeling a partner's latent state has a long lineage in Bayesian inverse planning~\cite{baker2009inverse} and Bayesian theory of mind~\cite{baker2017btom} with learned variants that predict agents from behavior~\cite{rabinowitz2018tomnet} and Bayesian delegation~\cite{wu2021toomanycooks}. Explicit belief representation in LLMs helps but is bounded by long-horizon context and hallucinated state~\cite{li2023tomllm}, and hypothesis-based agents generate and refine natural-language conjectures about partner strategy~\cite{hypotheticalminds}. Unlike belief estimators based on LLMs, we introduce a lightweight, non-LLM Bayesian method that infers which skill a teammate is currently executing from their publicly observable actions, to selectively interrupt the ego agent’s ongoing skill.
\paragraph{The belief-action gap in LLM agents.}
Our framing builds on the distinction between \emph{literal} Theory of Mind (predicting a partner) and \emph{functional} Theory of Mind (adapting behavior accordingly)~\cite{riemer2025tom}. Because LLM agents can fail to translate updated beliefs into action~\cite{xu2025saydo,dyntom2025}, we define the \emph{belief-action gap}: maintaining an accurate belief about a teammate’s role while continuing to execute an outdated skill. We introduce process-level metrics to measure this gap and evaluate whether belief-conditioned selective replanning reduces it.

\paragraph{Replanning and plan invalidation in hierarchical agents.}
LLM agents replan from task failure, self-reflection, or environment feedback~\cite{yao2023react,shinn2023reflexion} while more recent works trigger replanning for goal-state consistency or execution verification~\cite{kim2025reflact}. In these methods, replanning is primarily driven by task-level execution signals such as failed actions. We study a complementary setting in multi-agent coordination, where an executing skill may become obsolete before any task-level failure occurs because newly observed teammate behavior contradicts the current partner model. \textsc{BayesBeliefAgent} therefore uses these contradictions as the trigger for selective replanning.

\nocite{goel2023sociallight,zhang2025coordlight}

%% file: tex/method.tex
\section{Problem Formulation}
\label{sec:problem_formulation}

\textbf{Modeling.}
We model the problem as a hierarchical partially observable stochastic game in which agents act through temporally extended skills. 
The game is defined by the tuple $\langle \mathcal{N}, \mathcal{S}, \mathcal{A}, \mathcal{O}, \mathcal{T}, \Omega, R, \Psi, \boldsymbol{\pi}_{\mathrm{low}}, \boldsymbol{\beta}, \stateinitdist, \gamma \rangle$. Here, $\mathcal{N}=\{0,\ldots,n-1\}$ is the set of agents, $\mathcal{S}$ is the global state space, $\mathcal{A}=\mathcal{A}_0\times\cdots\times\mathcal{A}_{n-1}$ is the joint primitive-action space, and $\mathcal{O}=\mathcal{O}_0\times\cdots\times\mathcal{O}_{n-1}$ is the joint observation space. At time $t$, the environment is in state $s_t\in\mathcal{S}$, and agent $k$ receives observation $o_t^k\in\mathcal{O}_k$ sampled according to the observation model $o_t^k \sim \Omega^k(\cdot\mid s_t).$  Each agent executes a primitive action $a_t^k\in\mathcal{A}_k$, forming the joint action $\mathbf{a}_t=[a_t^0,\ldots,a_t^{n-1}]$. The transition model $\mathcal{T}$ determines the next state from $ s_{t+1}\sim\mathcal{T}(\cdot\mid s_t,\mathbf{a}_t)$, and the team receives the shared reward $R(s_t,\mathbf{a}_t)$. The finite skill library is denoted by $\Psi$, the collection of low-level action controllers by $\boldsymbol{\pi}_{\mathrm{low}}$, and the skill termination conditions by $\boldsymbol{\beta}$, as defined in the next paragraph. Finally, the initial state is sampled from $\stateinitdist$, and $\gamma\in[0,1]$ is the discount factor. The ego agent is denoted by $i$ and an unseen teammate by $j\neq i$.

\textbf{Temporally Extended Skills.} The skill library $\Psi$, where each $\psi\in\Psi$ represents a temporally extended task-level behavior.  The teammate selects its skill $\psi_t^j \sim \policyteam{\mathrm{high}}{j}\left( \cdot \mid \tau_t^j \right)$ using an unknown high-level policy $\policyteam{\mathrm{high}}{j}$ conditioned on its interaction history $\tau_t^j = \left(o_0^j, a_0^j, \ldots, o_{t-1}^j, a_{t-1}^j, o_t^j\right)$. Conditioned on its current skill $\psi_t^j$, the teammate executes a low level primitive action $a_t^j \sim \policyteam{\mathrm{low}}{j}\left(\cdot\mid o_t^j,\psi_t^j \right).$

While the ego agent observes the teammates' low-level actions $a_{t-1}^j$ as part of its own observations $o_t^i$, it does not know $\policyteam{\mathrm{high}}{j}$, or $\psi_t^j$. Therefore, the ego agent selects its skill $\psi_t^i \sim \policyego{\mathrm{high}}\left(\cdot \mid \tau_t^i,b_t(\psi_t^j) \right)$ using its high-level policy $\policyego{\mathrm{high}}$, conditioned on its trajectory $\trajt{t} = \left(o_0^i, a_0^i, \ldots, o_{t-1}^i, a_{t-1}^i, o_t^i\right)$  and belief $b_t$ (defined in the next section) over the teammate’s active skill. Conditioned on this skill, the ego agent's primitive action is $a_t^i
\sim \policyego{\mathrm{low}} \left(\cdot\mid o_t^i,\psi_t^i\right)$. Finally, each agent also has a termination rule $\beta^k$, which determines when its current skill ends. The collection of termination rules is denoted by $\boldsymbol{\beta}$. The termination function $\beta^k : \mathcal{O}_k \times \Psi \to [0,1]$ gives the probability that the skill terminates after time step $t$:
$\beta^k\left(o_{t+1}^k, \psi_t^k\right) \in [0,1]$.

\textbf{Online Partner-Skill Belief.}
Because the teammate’s active skill $\psi_t^j$ is not directly observable, the ego agent maintains a belief $b_t \in \Delta(\Psi)$, where $\Delta(\Psi)$ denotes the probability simplex over the skill library. For each $\psi \in \Psi$, the belief is defined as
\begin{equation}
  b_t(\psi) \,=\, \Pr\!\left(\psi_t^j = \psi \mid \tau_t^i\right).
  \label{eq:posterior}
\end{equation}
Since $o_t^i$ includes the teammate’s publicly observable primitive actions, $b_t$ summarizes which teammate skill best explains the evidence up to time $t$. The belief provides a probabilistic estimate of the teammate’s current task-level behavior, which the ego agent uses to choose a complementary skill and decide whether to interrupt and replan its own skill.

\textbf{Objective.}
The ego policy consists of a fixed hierarchical LLM planner, skill-conditioned low-level controllers, and an online partner-skill estimator. Therefore, the evaluation objective of the ego-agent's policy with the estimated partner's skill  is to maximize the total expected return:
\begin{equation}
\begin{aligned}
J(\policyego{\mathrm{high}})
&= \mathbb{E}_{\substack{
s_0 \sim \stateinitdist,\;
\mathbf{o}_t \sim \Omega(\cdot \mid s_t),\;
\mathbf{\psi}_t \sim \pi_{\mathrm{high}},\\
\mathbf{a}_t \sim \pi_{\mathrm{low}},\;
s_{t+1} \sim \mathcal{T}(\cdot \mid s_t,\mathbf{a}_t)}}
\!\left[\sum_{t=0}^{T} \gamma^t R(s_t,\mathbf{a}_t)\right].
\end{aligned}
\end{equation}



\begin{figure*}[t]
  \centering
  \includegraphics[width=0.8\textwidth]{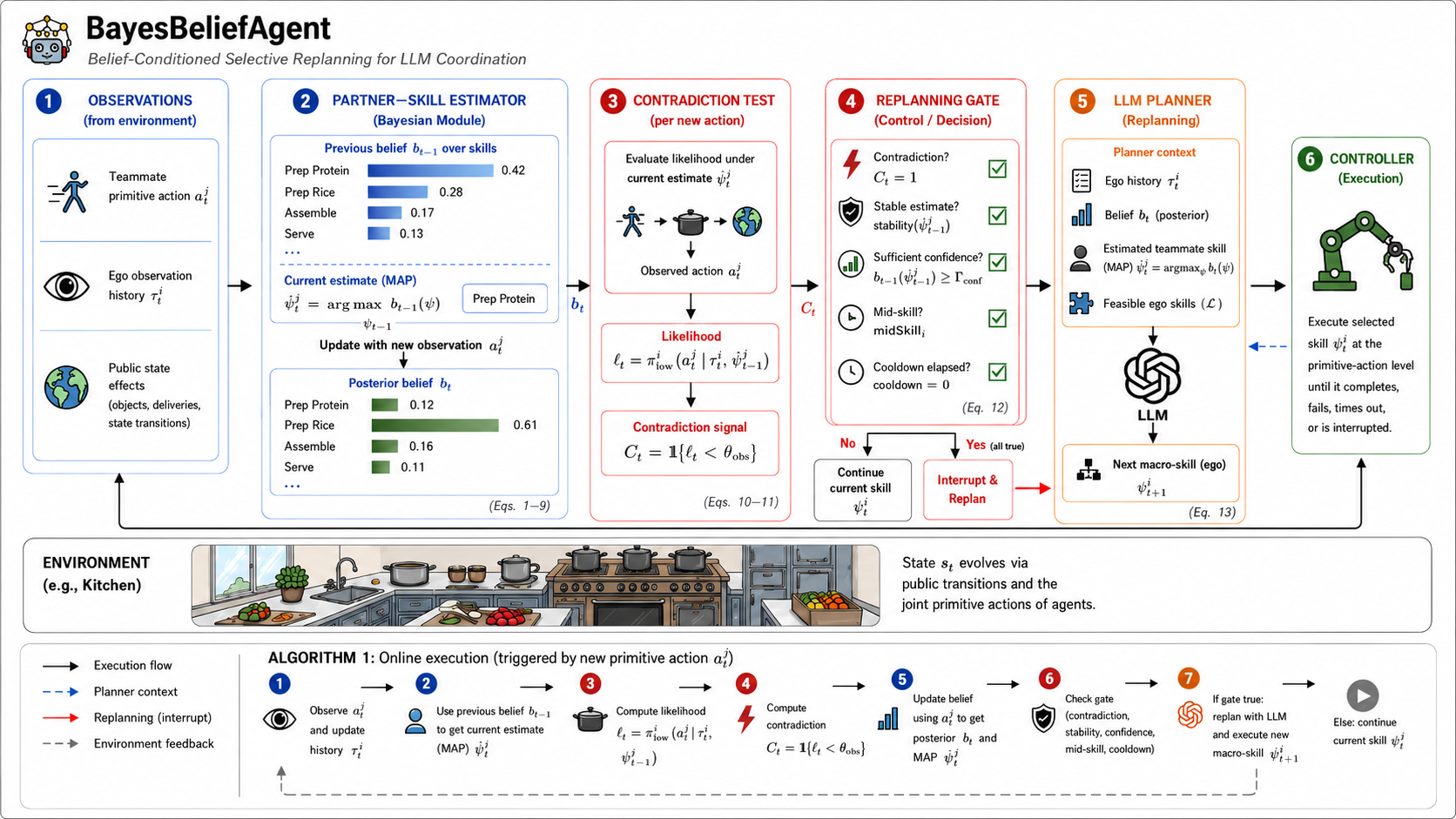}
\caption{\textbf{\textsc{BayesBeliefAgent} overview.} From the teammate's publicly observed primitive actions and the ego agent's observation history, the agent maintains a posterior over the teammate's active skill. Each newly observed teammate action is first evaluated under the previous MAP teammate-skill estimate. The posterior is then updated using the new observation. If the action is contradictory and the previous estimate is stable and sufficiently confident while the ego agent is mid-skill and out of cooldown, the current skill is interrupted and the LLM planner is queried using the updated posterior and MAP estimate. Otherwise, the agent continues executing its current skill.}
  \label{fig:overview}
\end{figure*}

\section{Method}
\label{method}

We present \textsc{BayesBeliefAgent}, a hierarchical ego agent that coordinates with an unseen teammate by maintaining an online posterior over the teammate's active skill. The ego agent observes the teammate's primitive actions and their effects on the public environment, but does not directly observe the skill being executed. We first estimate the teammate's active skill through a recursive Bayesian update. As described in Section~\ref{method:bayesianpartner}, the estimator tracks whether the teammate's skill may persist or change and then incorporates newly observed public evidence to obtain a posterior belief over the skill library. Once this posterior estimate is obtained, the ego agent uses it as coordination context for its high-level LLM planner. While an ego skill is already being executed, the agent may interrupt that skill and invoke the planner again when newly observed evidence is inconsistent with the current partner-skill estimate, as described in Section~\ref{method:skill_inteerupt}.

\subsection{Bayesian Partner-Skill Belief Update}
\label{method:bayesianpartner}

The ego agent maintains a recursive belief $b_t(\psi)$ over the teammate's active skill $\psi_t^j \in \Psi$, as introduced in Section~\ref{sec:problem_formulation}. Because the teammate's skill is not directly observable, the belief is updated from public evidence contained in the ego agent's observation history. We estimate the teammate's active skill through Bayesian inverse planning \cite{jin2024mmtom}.
The ego agent treats the teammate's skill $\psi_t^j\in\Psi$ as a latent variable and infers it from the primitive actions observed through its own interaction history. Let $
\tau_t^i = \left(
o_0^i,a_0^i,\ldots,o_{t-1}^i,a_{t-1}^i,o_t^i
\right)$
denote the ego agent's observation--action history. Because the teammate's publicly visible primitive action $a_{t-1}^j$ is contained in the ego observation $o_t^i$, this history provides the behavioral evidence used to infer the teammate's current skill.

For a candidate teammate skill $\psi\in\Psi$, the desired posterior is given in Eq.~\ref{eq:posterior}.
Applying Bayes' rule over a recent inference window from timestep $t'<t$, where the partner agent either completes its last skill or is a fixed window of $k$ timesteps
\begin{align}
\label{eq:skill_bayes_rule}
& P\bigl(\psi_t^j=\psi \mid \tau_t^i\bigr)
=\nonumber\\
&\frac{
P\bigl(\tau_{t'+1:t}^i \mid \psi_t^j=\psi,\tau_{t'}^i\bigr)
\, P\bigl(\psi_t^j=\psi \mid \tau_{t'}^i\bigr)
}{
\sum_{\psi'\in\Psi}
P\bigl(\tau_{t'+1:t}^i \mid \psi_t^j=\psi',\tau_{t'}^i\bigr)
\, P\bigl(\psi_t^j=\psi' \mid \tau_{t'}^i\bigr)
}
\nonumber\\
\propto
& P\bigl(\tau_{t'+1:t}^i \mid \psi_t^j=\psi,\tau_{t'}^i\bigr)\, b_{t'}(\psi),
\end{align}

where $\tau_{t'+1:t}^i$ denotes the new trajectory segment observed after the prior history $\tau_{t'}^i$, and $b_{t'}(\psi) \triangleq P\bigl(\psi_t^j=\psi \mid \tau_{t'}^i\bigr)$ is the prior belief over the teammate's active skill. In practice, we assume that the teammate's skill remains fixed within the inference window that is capped to $k$ steps.

\textbf{Initializing the Prior.} Therefore, the prior $b_{t'}$ is initialized from the observed history of completed partner skills. We use an add-$\alpha$ smoothed count model:
\begin{equation}
\label{eq:prior_init_counts}
 b_{t'}(\psi)
 \,=\,
 \frac{C_m\bigl(\psi\mid\tau_{t'}^i\bigr)+\alpha}{\sum_{\psi'\in\Psi}\Bigl(C_m\bigl(\psi'\mid\tau_{t'}^i\bigr)+\alpha\Bigr)}.
\end{equation}
where $\alpha>0$ is a smoothing hyperparameter. We use $\alpha=1$, corresponding to a uniform Dirichlet-style initialization. This smoothing assigns nonzero prior probability before the start of the current episode.

We update the episode-level counts after each inferred completed partner skill $\hat{\psi}_m^j$ at any completion timestep $m$ via
\begin{equation}
\label{eq:count_update}
C_m(\psi)
\,=\,
C_{m-1}(\psi)
+\mathds{1}\!\left[\hat{\psi}_m^j=\psi\right].
\end{equation}
The resulting prior provides a lightweight episode-level memory of the teammate's previously inferred behavior. Skills repeatedly supported in earlier segments receive greater prior probability, while the likelihood term in Eq.~\eqref{eq:skill_bayes_rule} can still rapidly shift the posterior when the teammate's newly observed actions are more consistent with a different skill.

\textbf{Simplifying the Belief Update.} The trajectory likelihood $P\left(
\tau_{t'+1:t}^i
\mid
\psi_t^j=\psi,\tau_{t'}^i
\right)$ in Eq.~\eqref{eq:skill_bayes_rule} contains both the teammate's primitive actions from its action policy  $\policyteam{\mathrm{low}}{j}$ and the resulting observations and can be factorized as
\begin{align}
\prod_{l=t'+1}^{t}
\policyteam{\mathrm{low}}{j}\left(
a_l^j
\mid
o_l^j,\psi
\right)
P\left(
a_l^i
\mid
\tau_l^i
\right)
P\left(
o_{l+1}^i
\mid
\tau_l^i,a_l^i,a_l^j
\right).\nonumber\
\label{eq:trajectory_likelihood_factorization}
\end{align}

Therefore,  the partner-skill belief update in Eq.\ref{eq:skill_bayes_rule} can be written as  
\begin{equation}
b_t(\psi) =
\eta
b_{t'}(\psi)
\prod_{l=t'+1}^{t}
\policyteam{\mathrm{low}}{j}
\left(
a_l^j
\mid
o_l^j,\psi
\right),
\label{eq:skill_posterior_true_controller}
\end{equation}
where the normalization constant is
\begin{equation}
\eta^{-1} =
\sum_{\psi'\in\Psi}
b_{t'}(\psi')
\prod_{l=t'+1}^{t}
\policyteam{\mathrm{low}}{j}
\left(
a_l^j
\mid
o_l^j,\psi'
\right).
\label{eq:skill_posterior_normalizer}
\end{equation}

In practice, the true teammate observation $o_l^j$ and low-level controller
$\policyteam{\mathrm{low}}{j}$ are not available to the ego agent. We therefore approximate the action likelihood using the ego agent's own observations of its partner from its own history $\tau_l^i$ and its low-level action controller $\policyego{\mathrm{low}}$ as 
\begin{equation}
\policyteam{\mathrm{low}}{j}
\left(
a_l^j
\mid
o_l^j,\psi
\right)
\approx
\policyego{\mathrm{low}}
\left(
a_l^j
\mid
\tau_l^i,\psi
\right).
\label{eq:approximate_teammate_controller}
\end{equation}


\subsection{Belief-Guided Skill Interruption}
\label{method:skill_inteerupt}

The posterior belief over the teammate’s skill is used both as context for the high-level planner and to decide when the ego agent should interrupt its current skill. 

First, the ego agent obtains a point estimate of the teammate's active skill $\hat{\psi}_{t-1}^j$ from the previous belief:
\begin{equation}
\label{eq:current_partner_skill_estimate}
\hat{\psi}_{t-1}^j
\;=\;
\arg\max_{\psi\in\Psi} \, b_{t-1}(\psi).
\end{equation}

Then, the ego agent evaluates the likelihood of teammate actions under this estimated skill. Because the teammate's true low-level controller and observation are unavailable, we use the same approximate action model introduced in Eq.~\eqref{eq:approximate_teammate_controller}:
\begin{equation}
\ell_t
=
\prod_{l=t'+1}^{t-1} 
\policyego{\mathrm{low}}
\left(
a_l^j
\mid
\tau_t^l,\hat{\psi}_{t-1}^j
\right).
\label{eq:current_skill_likelihood}
\end{equation}
Finally, this likelihood measures how well the current partner-skill estimate explains the teammate's newly observed behavior to determine whether to interrupt the ego agent's ongoing skill. A low value indicates that the teammate's action is inconsistent with the behavior expected under $\hat{\psi}_{t-1}^j$.

We define the resulting contradiction signal as
\begin{equation}
C_t
=
\mathds{1}
\left[
\ell_t < \theta_{\mathrm{obs}}
\right],
\label{eq:partner_skill_contradiction}
\end{equation}
where $\theta_{\mathrm{obs}}$ is a likelihood threshold. The contradiction test is evaluated before the new action is incorporated into the posterior in Eq. \ref{eq:skill_posterior_true_controller}. This ordering is important because, after the belief update, the posterior may shift toward a different skill that better explains $a_t^j$. Evaluating the likelihood before this update allows the ego agent to detect when its previous partner-skill estimate has been contradicted by new evidence.

\textbf{Gated Skill Interruption.}
A single unlikely action may result from noise or temporary motion rather than a meaningful change in the teammate's skill. We therefore combine the contradiction signal with additional conditions before interrupting the ego agent's current skill:
\begin{equation}
\begin{aligned}
\mathrm{Replan}_t
=
\mathds{1}\bigg[
&
C_t = 1
\;\wedge\;
\mathrm{stable}\left(\hat{\psi}_{t-1}^j\right)
\\
&\wedge\;
b_{t-1}\left(\hat{\psi}_{t-1}^j\right)
\geq
\Gamma_{\mathrm{conf}}
\\
&\wedge\;
\mathrm{midSkill}_t
\;\wedge\;
\mathrm{cooldown}_t = 0
\bigg].
\end{aligned}
\label{eq:replan}
\end{equation}
Here, $\mathrm{stable}\left(\hat{\psi}_{t-1}^j\right)$ requires the MAP skill estimate to have remained unchanged over a fixed number of recent belief updates. This prevents the agent from interrupting its skill while the partner estimate is still changing. The confidence threshold $\Gamma_{\mathrm{conf}}$ requires the current estimate to have sufficient posterior probability. The condition $\mathrm{midSkill}_t$ ensures that interruption is only considered while the ego agent is executing a skill, and $\mathrm{cooldown}_t$ prevents repeated replanning over a short period. We present the sensitivity of the planner to these gates and outline hyperparameters in the Appendix.

To summarize, the ego agent replans only when a newly observed teammate action is unlikely under a stable, confident partner-skill estimate. Therefore, the mechanism does not need the complementary ego skill; it detects when the teammate's behavior is inconsistent with the stable, committed skill estimate held before the new observation, and uses that contradiction to trigger replanning.

\subsection{Replanning}
\label{sec:method-replan}

When Eq.~\eqref{eq:replan} is satisfied, the ego agent interrupts its current skill and invokes the high-level LLM planner. The planner receives the current observation history, the updated posterior belief $b_t$, the resulting MAP teammate-skill estimate
\begin{equation}
\hat{\psi}_t^j
=
\arg\max_{\psi\in\Psi}
b_t(\psi),
\end{equation}
and the set of currently feasible ego skills. The planner $\policyego{\mathrm{high}}$, which is an LLM selects a new ego skill. Consequently, the partner-skill belief affects planning in two ways. First, the posterior and its MAP estimate are provided to the LLM planner as coordination context. Second, the likelihood of newly observed behavior under the current skill estimate determines whether the planner should be invoked while an ego skill is already in progress. This allows the agent to continue executing its current skill when the teammate behaves as expected, while still responding when new evidence contradicts the current partner-skill estimate. We provide the complete algorithm in the Appendix.

%% file: tex/experiments.tex
\section{Experiments}
\label{sec:experiments}

Our experiments answer the following research questions:
\begin{enumerate}[label=\textbf{RQ\arabic*.}, leftmargin=*, itemsep=2pt]
\item \textbf{Coordination Performance.}
How does \textsc{BayesBeliefAgent} perform when paired with diverse, preference-specialized partners (Sec. \ref{sec:rq1_benchmark})?

\item \textbf{Literal--Functional ToM Gap.}
Does solely maintaining an accurate belief about a teammate's skill suffice to prevent coordination failures (Sec. \ref{sec:rq2})?

\item \textbf{Posterior-Guided Control Ablation.}
Does only using the teammate posterior as a \emph{control signal} to interrupt an executing skill close the belief-action gap more effectively than using the same posterior only as \emph{context} in the planner prompt (Sec. \ref{sec:rq3})?

\item \textbf{Selective Replanning.}
Is contradiction-based interruption more efficient at improving coordination than generic replanning heuristics (Sec. \ref{sec:rq4})?
\end{enumerate}

\subsection{Setup}
\label{sec:setup}

\paragraph{Environments and partners.}
We evaluate in the Overcooked/Burrito domain across three layouts (Open, Ring, and Forced Coordination (FC)), and three partner groups, yielding nine settings. The evaluation population contains 12 Behavior Preference (BP) agents~\citep{wang2024zsc} that favor distinct task subgoals through reward shaping. We divide them into three groups of four: \textbf{Logistics/G1} (Plating and Washing), \textbf{Cooking/G2} (Chopping and Potting), and \textbf{Sourcing/G3} (Mushroom and Rice dispensing). Within each group, the four agents correspond to the $2^2$ combinations of their two preferences. Rewards for Delivering Dishes $(+20)$, Grilling Meat $(+20)$, and Taking Meat $(+10)$ are fixed across all partners to maintain task competence while preserving behavioral diversity. Each reported layout-group result aggregates five seeds across four partners. Full partner-generation details are provided in the Appendix.

\paragraph{Baselines and conditions.} 
We compare \textsc{BayesBeliefAgent} against \textsc{ProAgent}~\citep{zhang2024proagent}, a hierarchical LLM planner without online belief revision; Hypothetical Minds~\citep{hypotheticalminds}, a hypothesis-based partner-modeling method; \textsc{GAMMA}~\citep{liang2024learningcooperatehumansusing}, an RL-based ad hoc teaming agent; and \textsc{TALENTS}~\citep{li2025adaptively}, a zero-shot coordination baseline. All LLM-based conditions use either GPT-4o or GPT-5.2.


\begin{table*}[htbp]
\centering
\small
\caption{Coordination performance (mean reward $\pm$ std.\ dev.) across all three BP partner groups. FC = Forced Coordination. Best LLM-based result per cell in \textbf{bold}; best overall underlined. \textsc{BayesBeliefAgent} outperforms all LLM-based agents.}
\label{tab:results_all_groups}
\resizebox{\linewidth}{!}{%
\begin{tabular}{lcccccccccc}
\toprule
& \multicolumn{3}{c}{\textbf{Group 1 (Logistics)}} & \multicolumn{3}{c}{\textbf{Group 2 (Cooking)}} & \multicolumn{3}{c}{\textbf{Group 3 (Sourcing)}} \\
\cmidrule(lr){2-4}\cmidrule(lr){5-7}\cmidrule(lr){8-10}
\textbf{Agent} & \textbf{FC} & \textbf{Ring} & \textbf{Open}
               & \textbf{FC} & \textbf{Ring} & \textbf{Open}
               & \textbf{FC} & \textbf{Ring} & \textbf{Open} \\
\midrule
BP        & \underline{95} \tiny{(116)}  & 927 \tiny{(406)} & \underline{1625} \tiny{(457)}
                      & \underline{84} \tiny{(159)} & 622 \tiny{(246)} & \underline{484} \tiny{(179)}
                      & \underline{76} \tiny{(130)} & 793 \tiny{(228)} & 792 \tiny{(334)} \\
GAMMA                 & 46 \tiny{(79)}  & \underline{1124} \tiny{(225)} & 1189 \tiny{(415)}
                      & 64 \tiny{(111)}  & 649 \tiny{(159)} & 360 \tiny{(114)}
                      & 54 \tiny{(77)}  & 864 \tiny{(216)} & 692 \tiny{(167)} \\
TALENTS               & 53 \tiny{(91)}  & 1019 \tiny{(307)} & 1212 \tiny{(193)}
                      & 34 \tiny{(60)}  & \underline{935} \tiny{(298)} & 398 \tiny{(158)}
                      & 38 \tiny{(49)}  & \underline{1237} \tiny{(308)} & 1101 \tiny{(321)} \\
\midrule
Hypothetical-Minds  &11 \tiny{(19)} & 388 \tiny{(361)} & 608 \tiny{(274)} & 0 \tiny{(0)} & 188 \tiny{(64)} & 369 \tiny{(164)} & 0 \tiny{0} & 275 \tiny{(124)} & 480 \tiny{(239)} \\
ProAgent              & 0 \tiny{(0)}    & 313 \tiny{(252)} & 732 \tiny{(310)}
                      & 0 \tiny{(0)}    & 161 \tiny{(143)} & 388 \tiny{(253)}
                      & 9 \tiny{(15)}  & 271 \tiny{(234)} & 549 \tiny{(327)} \\
ProAgent (GPT-5.2)    & 0 \tiny{(0)}    & 200 \tiny{(182)} & 629 \tiny{(252)}
                      & 0 \tiny{(0)}    & 172 \tiny{(80)}  & 392 \tiny{(140)}
                      & 9 \tiny{(16)}  & 337 \tiny{(239)} & 569 \tiny{(257)} \\
Ours (GPT-4o)         & \textbf{14} \tiny{(21)}    & 663 \tiny{(468)} & \textbf{1533} \tiny{(410)}
                      & \textbf{9} \tiny{(16)}    & 231 \tiny{(232)} & 337 \tiny{(270)}
                      & 18 \tiny{(25)}                  & \textbf{506} \tiny{(161)}                  & \textbf{\underline{1420}} \tiny{(390)} \\
Ours (GPT-5.2)        & 0 \tiny{(0)}    & \textbf{705} \tiny{(302)} & 1485 \tiny{(340)}
                      & \textbf{9} \tiny{(16)} & \textbf{285} \tiny{(114)} & \textbf{395} \tiny{(234)}
                      & \textbf{45} \tiny{(40)} & 470 \tiny{(173)} & 1360 \tiny{(306)} \\
\bottomrule
\end{tabular}}
\end{table*}

\paragraph{Metrics.}
We report episodic team rewards as measures of task performance. To evaluate \emph{literal} Theory of Mind, we additionally report teammate-skill prediction accuracy. To evaluate \emph{functional} Theory of Mind, we report the complementary-skill rate within three steps of a detected contradiction (\emph{Comp@3}), and the duplicate-skill rate.  The ground truth complementary skill for coordination scenarios is determined by humans (See Appendix). 

Our primary diagnostic metric is the \emph{belief-action gap rate}. At each high-level decision point, we first determine whether the agent's committed estimate of the teammate's skill is correct, or satisfies a specified confidence criterion. Among these points, the belief-action gap rate measures the fraction for which the ego agent nevertheless selects or continues a stale, duplicate, or non-complementary macro-action. Thus, this metric assesses whether that belief is translated into appropriate control. A lower rate indicates a smaller divergence between literal and functional Theory of Mind. Additionally, we measure the replan rate per episode to quantify the computational cost of each interruption strategy. 

\subsection{RQ1: Benchmarking Coordination}
\label{sec:rq1_benchmark}

Table~\ref{tab:results_all_groups} reports coordination performance across the three partner clusters and three layouts. \textsc{BayesBeliefAgent} outperforms \textsc{ProAgent} and \textsc{Hypothetical Minds} in every layout. In the \textbf{Open} and \textbf{Ring} layouts, \textsc{BayesBeliefAgent} improves performance relative to \textsc{ProAgent}, with the largest gains against Group~1 and Group~3 partners. Qualitatively, these improvements arise because \textsc{BayesBeliefAgent} anticipates large partner skill switches (e.g., to plating or washing in Group~1, or mushroom dispensing in Group~3) and selects a complementary skill in response. However, the RL-based baselines, particularly \textsc{TALENTS} and \textsc{GAMMA}, remain stronger than the LLM-based methods in most environments. This gap suggests that, although online teammate-belief adaptation improves coordination via the hierarchical LLM agent, the underlying LLM-based high-level skill planner remains a major performance bottleneck and requires substantial improvement. Improving the underlying high-level LLM planner is an important direction that we leave to future work.

\subsection{RQ2: Literal-Functional ToM Gap}
\label{sec:rq2}

\begin{table}[htbp]
\centering
\small
\caption{Mechanism metrics averaged across partner groups G1--G3. Despite nearly identical teammate-skill prediction accuracy, \textsc{BayesBeliefAgent} reduces the belief--action gap and duplicate-skill rate and improves contradiction-conditioned recovery relative to \textsc{ProAgent}, with the largest gains in Open and Ring. Arrows indicate the improving direction.}
\label{tab:rq2_results}
\begin{tabular}{llrrr}
\toprule
Metric & Method & Open & Ring & FC \\
\midrule
\multirow{2}{*}{Skill acc.\ ($\uparrow$)}
& \textsc{BayesBeliefAgent} & 0.79 & 0.71 & 0.61 \\
& \textsc{ProAgent} & 0.78 & 0.71 & 0.60 \\
\midrule
\multirow{2}{*}{Gap rate ($\downarrow$)}
& \textsc{BayesBeliefAgent} & 0.20 & 0.28 & 0.42 \\
& \textsc{ProAgent} & 0.41 & 0.38 & 0.47 \\
\midrule
\multirow{2}{*}{Comp@3 ($\uparrow$)}
& \textsc{BayesBeliefAgent} & 0.66 & 0.54 & 0.31 \\
& \textsc{ProAgent} & 0.39 & 0.43 & 0.26 \\
\midrule
\multirow{2}{*}{Dup.\ skill ($\downarrow$)} 
& \textsc{BayesBeliefAgent} & 0.15 & 0.23 & 0.37 \\
& \textsc{ProAgent} & 0.30 & 0.30 & 0.41 \\
\bottomrule
\end{tabular}
\end{table}

Table~\ref{tab:rq2_results} shows that \textsc{BayesBeliefAgent} (GPT-4o) and  ProAgent achieve nearly identical predicted teammate-skill accuracy across all layouts, yet differ substantially in functional coordination. The posterior-gated interruption in \textsc{BayesBeliefAgent} reduces the belief-action gap from $0.41$ to $0.20$ in Open and from $0.38$ to $0.28$ in Ring, while also improving Comp@3 and reducing duplicate-skill behavior. The gains are smaller in Forced Coordination, where limitations of the underlying planner remain more pronounced. These results reveal a literal-functional Theory-of-Mind gap: accurately inferring the teammate's skill is insufficient if the ego agent remains committed to a skill selected under an outdated belief. Thus, \textsc{BayesBeliefAgent} addresses a key coordination bottleneck not by further improving belief accuracy, but by translating updated teammate beliefs into timely control decisions to bridge the literal-functional ToM gap.


\subsection{RQ3: Posterior-Guided Control Ablation}
\label{sec:rq3}

We next examine whether the teammate's skill posterior is more effective as planner context or as a control signal. We present four ablations in Table~\ref{tab:ablation} that test posterior prompting and posterior-gated interruption: \textsc{Full} (equivalent to \textsc{BayesBeliefAgent}) uses both pathways, \textsc{No replan} uses only prompting, \textsc{No prompt} uses only posterior-guided interruption, and \textsc{No belief} uses neither. Thus, the comparison between \textsc{No prompt} and \textsc{No replan} contrasts control-only and context-only use of the same posterior, while \textsc{Full} and \textsc{No replan} isolate the additional effect of interruption when posterior context is held fixed.

 \begin{table}[htbp]
\centering
\scriptsize
\caption{We report the mean reward $\pm$ standard deviation across posterior-guided control ablations. Posterior-gated interruption generally outperforms posterior prompting alone, indicating its effectiveness as a control signal.}
\label{tab:ablation}
\begin{tabular}{lrrrr}
\toprule
Setting & Full & No prompt & No replan & No belief \\
\midrule
Open G1 & $1533{\pm}\tiny{410}$ & $1405{\pm}\tiny{430}$ & $1190{\pm}\tiny{455}$ & $1080{\pm}\tiny{470}$ \\
Open G2 & $337{\pm}\tiny{270}$ & $305{\pm}\tiny{250}$ & $263{\pm}\tiny{152}$ & $274{\pm}\tiny{290}$ \\
Open G3 & $1420{\pm}\tiny{390}$ & $1325{\pm}\tiny{420}$ & $1110{\pm}\tiny{440}$ & $1015{\pm}\tiny{465}$ \\
Ring G1 & $663{\pm}\tiny{468}$ & $610{\pm}\tiny{430}$ & $566{\pm}\tiny{348}$ & $212{\pm}\tiny{206}$ \\
Ring G2 & $231{\pm}\tiny{232}$ & $225{\pm}\tiny{215}$ & $264{\pm}\tiny{218}$ & $215{\pm}\tiny{227}$ \\
Ring G3 & $506{\pm}\tiny{161}$ & $475{\pm}\tiny{185}$ & $439{\pm}\tiny{177}$ & $607{\pm}\tiny{582}$ \\
FC G1 & $14{\pm}\tiny{21}$ & $13{\pm}\tiny{19}$ & $11{\pm}\tiny{18}$ & $9{\pm}\tiny{17}$ \\
FC G2 & $9{\pm}\tiny{16}$ & $8{\pm}\tiny{15}$ & $8{\pm}\tiny{14}$ & $7{\pm}\tiny{13}$ \\
FC G3 & $18{\pm}\tiny{25}$ & $17{\pm}\tiny{24}$ & $15{\pm}\tiny{22}$ & $13{\pm}\tiny{21}$ \\
\bottomrule
\end{tabular}
\end{table}

Table~\ref{tab:ablation} shows that \textsc{No prompt} generally outperforms \textsc{No replan}. This indicates that the critical contribution of the posterior is not merely to inform the planner, but to trigger replanning when the current skill becomes inconsistent with updated teammate behavior. Even without posterior context in the prompt, posterior-gated interruption enables better coordination than providing the same posterior only as planner context. The comparison between \textsc{Full} and \textsc{No replan} further supports this conclusion. Moreover, the added interruption while holding posterior prompting fixed reduces the belief--action gap in five of seven settings (see Appendix). Overall, RQ3 shows that directly coupling teammate beliefs to replanning is more effective than relying on the planner to translate those beliefs into control decisions.

\subsection{RQ4: Selective Replanning}
\label{sec:rq4}

A natural alternative explanation is that coordination improves simply because the agent replans more often, rather than because behavioral surprise identifies informative decision points for interruption. We compare contradiction-based interruption against three generic replanning heuristics. \textsc{Periodic-10} interrupts and replans every ten environment steps. \textsc{Completion/held-item} triggers replanning whenever the partner completes a skill or is observed holding a task-relevant item. \textsc{LLM-judge} uses a separate LLM-based decision module that observes the partner's public behavior and determines whether the ego agent should replan.

Table~\ref{tab:triggers} shows that the generic triggers can attain rewards comparable to \textsc{Full}, but only by replanning far more frequently. On Open~G1, \textsc{Periodic-10}, \textsc{Completion/held-item}, and \textsc{LLM-judge} invoke $43$, $169$, and $119$ replans per episode, respectively, compared with only $2.0$ for \textsc{BayesBeliefAgent}, while achieving similar rewards ($1490$, $1510$, and $1480$ versus $1533$). This pattern holds broadly across settings: the generic heuristics typically trigger tens to more than one hundred replans per episode, whereas contradiction-based interruption operates with low single-digit replan counts in most cases. Although some generic triggers achieve higher numerical reward in individual settings, none does so consistently across environments, and these gains come at substantially greater replanning cost. These results indicate that the benefit of contradiction-based interruption is not simply more frequent replanning, but greater selectivity in identifying when an executing skill should be revised.


\begin{table}[htbp]
\centering
\scriptsize
\caption{We report the mean reward / replans per episode to compare replanning triggers. All LLM-planner conditions receive identical public observations and belief prompt; they differ only in the interruption signal. \textsc{BayesBeliefAgent} (Full) attains comparable reward with far fewer replans than the heuristic or LLM-based triggers.}
\label{tab:triggers}
\begin{tabular}{lrrrrrrr}
\toprule
Setting & Full & Periodic-10 & Compl./held & LLM-judge  \\
\midrule
Open G1 & 1533 / 2.0 & 1490 / 43.0 & 1510 / 169.0 & 1480 / 119.0  \\
Open G2 & 337 / 1.8 & 515 / 15.0 & 304 / 50.4 & 403 / 34.0  \\
Open G3 & 1420 / 3.0 & 1395 / 51.0 & 1370 / 115.0 & 1400 / 98.0  \\
Ring G1 & 663 / 1.6 & 695 / 66.9 & 629 / 107.1 & 721 / 97.1 \\
Ring G2 & 231 / 7.5 & 218 / 76.3 & 202 / 87.5 & 201 / 110.6 \\
Ring G3 & 506 / 9.3 & 416 / 52.8 & 455 / 110.9 & 558 / 103.3\\
FC G1 & 14 / 0.5 & 12 / 78.0 & 16 / 8.0 & 11 / 92.0  \\
FC G2 & 9 / 0.3 & 8 / 81.0 & 10 / 6.0 & 7 / 96.0\\
FC G3 & 18 / 0.6 & 15 / 76.0 & 19 / 10.0 & 14 / 89.0\\
\bottomrule
\end{tabular}
\end{table}



%% file: tex/appendix.tex
\section*{Appendix}

\section{Full Algorithm Specification}
\label{app:algorithm}

Algorithm~\ref{alg:bayes-belief-full} gives the complete per-step
operation of \textsc{BayesBeliefAgent}. To keep the pseudocode compact,
we introduce the following notation. Let
\[
L_t(\psi)
\triangleq
\policyego{\mathrm{low}}
\left(
a_l^j
\mid
\tau_l^i,\psi
\right).
\]
denote the implemented action likelihood corresponding to the
approximate teammate controller in
Eq.~\eqref{eq:approximate_teammate_controller}.
The operator $\textsc{Prior}(N_m)$ returns the smoothed count-based
distribution in Eq.~\eqref{eq:prior_init_counts}, where the algorithmic
count array $N_m$ corresponds to $C_m$ in the method.

We write the normalized one-step Bayesian correction as
\[
\textsc{Correct}(b,l_t)(\psi)
=
\frac{
    l_t(\psi)b(\psi)
}{
    \sum_{\psi'\in\Psi}
    l_t(\psi')b(\psi')
},
\]

where $l_t$ is defined in Eq.~\eqref{eq:current_skill_likelihood}.
Applying this correction recursively within a partner-skill segment
accumulates the likelihood product in
Eqs.~\eqref{eq:skill_posterior_true_controller}--%
\eqref{eq:skill_posterior_normalizer}, using the approximation in
Eq.~\eqref{eq:approximate_teammate_controller}.

The operator $\textsc{Gate}$ implements Eq.~\eqref{eq:replan}, with
\[
\mathrm{stable}(\hat\psi_{t-1}^j)
\equiv [u\geq K]
\]
and 
\[
\qquad
\mathrm{midSkill}_t
\equiv
[\mu\neq\textsc{null}\wedge\neg\mathrm{complete}(\mu)],
\]
and $\mathrm{cooldown}_t=0$ represented by $c=0$.
Finally, $\textsc{Plan}$ constructs the planner prompt, queries the
high-level LLM, and parses a skill from the currently feasible
macro-skill set.

\begin{algorithm}[t]
\small
\caption{\textsc{BayesBeliefAgent}: per-step belief update and selective replanning}
\label{alg:bayes-belief-full}
\begin{algorithmic}[1]
\Require partner-skill set $\Psi$, ego macro-skill set $\mathcal{M}$,
planner $\policyego{\mathrm{high}}$, controller $\policyego{\mathrm{low}}$
\Require smoothing $\alpha$, thresholds
$\theta_{\mathrm{obs}}$ and $\Gamma_{\mathrm{conf}}$,
stability window $K$, inference window $k$,
cooldown duration $C_{\mathrm{cool}}$

\State $N_0(\psi)\gets0$ for all $\psi\in\Psi$;
       $b_0\gets\Call{Prior}{N_0}$
\State $r\gets1$; $\mu\gets\textsc{null}$; $c,u\gets0$;
       $\hat\psi^j_0,\psi^j_{-1}\gets\arg\max_{\psi\in\Psi}b_0(\psi)$

\For{$t=1,2,\ldots$ until the episode terminates}
    \State Observe $o^i_t$ and infer $\hat a_{t-1}^{j}$ from
           $o^i_t$

    \State $\hat\psi^{j}_{t-1}\gets
           \arg\max_{\psi\in\Psi}b_{t-1}(\psi)$

    \State
           $u\gets
           1+\mathds{1}[\hat\psi^j_{t-1}=\hat\psi^j_{t-2}]u$

    \State $\ell_t\gets
           \displaystyle\prod_{l=\max\{r,t-k\}}^{t-1}
           L_l(\hat\psi^j_t)$

    \State
           $C_t\gets\mathds{1}[\ell_t<\theta_{\mathrm{obs}}]$

    \State $\mathrm{Replan}_t\gets
           \Call{Gate}{
               C_t,u,b_{t-1}(\hat\psi^j_{t-1}),\mu,c
           }$

    \State $\tilde b_t\gets
           \Call{Correct}{b_{t-1},l_t}$

    \If{the teammate visibly completes a skill $\psi_t^{\mathrm{comp}}$}

        \State $N_t(\psi)\gets
               N_{t-1}(\psi)
               +\mathds{1}[\hat\psi_t^{\mathrm{comp}}=\psi]$
               for all $\psi\in\Psi$

        \State $b_t\gets\Call{Prior}{N_t}$;
               $r\gets t$; $u\gets0$
    \Else
        \State $b_t\gets\tilde b_t$
    \EndIf

    \State \textbf{if}
           $\mathrm{complete}(\mu)\vee\mathrm{Replan}_t$
           \textbf{then} $\mu\gets\textsc{null}$

    \State \textbf{if} $\mathrm{Replan}_t$
           \textbf{then} $c\gets C_{\mathrm{cool}}$
           \textbf{else} $c\gets\max(c-1,0)$

    \If{$\mu=\textsc{null}$}
        \State $\hat\psi_t^j\gets
               \arg\max_{\psi\in\Psi}b_t(\psi)$

        \State $\mu\gets
               \Call{Plan}{
                   o^i_t,b_t,\hat\psi_t^j,
                   \mathcal{M}_{\mathrm{feasible}}(s_t)
               }$
    \EndIf

    \State Execute one low-level controller step conditioned on $\mu$
\EndFor
\end{algorithmic}
\end{algorithm}

\subsection{Algorithm Walkthrough}
\label{app:algorithm:explanation}

Lines~1--2 initialize the episode-level completed-skill counts, the
smoothed prior in Eq.~\eqref{eq:prior_init_counts}, the current
partner-skill segment, the ego macro-skill, the cooldown, and the
stable-MAP counter. Lines~3--4 begin the per-step loop, observe the
public state, and infer the teammate's latest primitive action from
consecutive observations. Lines~5--7 compute the pre-correction MAP
skill estimate, update its stability counter, evaluate the recent
action likelihood in Eq.~\eqref{eq:current_skill_likelihood}, form the
contradiction signal in Eq.~\eqref{eq:partner_skill_contradiction}, and
apply the gated interruption rule in Eq.~\eqref{eq:replan}. Line~8 then
incorporates the newly observed action through the Bayesian correction
defined by
Eqs.~\eqref{eq:skill_posterior_true_controller}--%
\eqref{eq:skill_posterior_normalizer}, using the approximate controller
in Eq.~\eqref{eq:approximate_teammate_controller}. Lines~9--15 handle
partner-skill boundaries: when a visible completion occurs, the agent
infers the completed skill, updates its count using
Eq.~\eqref{eq:count_update}, reinitializes the prior, and begins a new
segment; otherwise, it retains the corrected within-segment posterior.
Lines~16--17 clear the ego macro-skill after either ordinary completion
or contradiction-triggered interruption and update the replanning
cooldown. Finally, Lines~18--21 query the high-level planner whenever no
macro-skill is active, using the updated posterior, its MAP estimate,
and the currently feasible skills, while Lines~22--23 execute one
low-level controller step and continue to the next timestep.

\section{Partner Population and Baseline Implementations}
\label{app:implementation}

\subsection{Behavior-Preferring Partner Population}
\label{app:implementation:bp-population}
We train a population of 12 behavior-preferring (BP) partner
policies. Following the event-based reward construction of
\citet{wang2024zsc}, each BP policy is optimized using a shaped
reward
\begin{equation}
    r^{\mathrm{BP}}_{\mathbf{w}}(s_t,\mathbf{a}_t)
    =
    r_{\mathrm{task}}(s_t,\mathbf{a}_t)
    +
    \sum_{f \in \mathcal{F}}
    w_f \phi_f(s_t,\mathbf{a}_t),
    \label{eq:bp-reward}
\end{equation}
where $r_{\mathrm{task}}$ is the original team reward,
$\phi_f(s_t,\mathbf{a}_t)$ indicates whether behavior feature $f$
occurred at step $t$, and $w_f$ is its shaping weight. Retaining the
task reward encourages the BP policy to express its assigned
preference while continuing to contribute to task completion.

The population is divided into three groups representing logistics,
cooking, and ingredient-sourcing preferences. Within each group, two
behavior features have two possible reward weights and a third feature
has a fixed positive weight. We train all four combinations of the two
variable weights, producing four BP policies per group and twelve
policies in total. All twelve policies use the same PPO architecture
and optimization procedure, described in
Appendix~\ref{app:implementation:bp-training}.

\subsection{Behavior-Preferring Reward Configurations}
\label{app:implementation:bp-rewards}

\label{app:implementation:bp-rewards}

Table~\ref{tab:bp_rewards_final} specifies the event-weight space used
for each BP group. A set such as $\{-30,+20\}$ denotes two alternative
weights. Within each group, the two non-fixed weight sets are crossed
to obtain four reward configurations.

\begin{table}[htbp]
\centering
\small
\caption{Event-weight configurations used to train the 12 BP partner
policies. Within each group, the two non-fixed weights are crossed,
producing four policies; the third weight is fixed across those four
policies.}
\label{tab:bp_rewards_final}
\begin{tabular}{llc}
\toprule
\textbf{Group} & \textbf{Behavior Feature}
& \textbf{Reward(s)} \\
\midrule
\multirow{3}{*}{Group 1 (Logistics)}
    & Plating Ingredients       & $\{-30,+20\}$ \\
    & Washing Plates            & $\{-30,+20\}$ \\
    & Delivering Dishes         & $+20$ (fixed) \\
\midrule
\multirow{3}{*}{Group 2 (Cooking)}
    & Chopping Ingredients      & $\{-30,+20\}$ \\
    & Potting Rice              & $\{-30,+20\}$ \\
    & Grilling Meat             & $+20$ (fixed) \\
\midrule
\multirow{3}{*}{Group 3 (Sourcing)}
    & Taking Mushroom           & $\{-15,+10\}$ \\
    & Taking Rice               & $\{-15,+10\}$ \\
    & Taking Meat               & $+10$ (fixed) \\
\bottomrule
\end{tabular}
\end{table}

\subsection{Behavior-Preferring Policy Training}
\label{app:implementation:bp-training}

\label{app:implementation:bp-training}

Each reward configuration is trained independently using RLlib PPO.
All policies use the same observation representation, policy network,
optimizer, and training schedule. We train each policy for
3 million environment steps using a learning rate of
\texttt{5e-5}, discount factor $\gamma=\texttt{0.999}$, GAE
parameter $\lambda=\texttt{0.999}$, PPO clipping parameter
$\epsilon=0.2$, train batch size 8192, minibatch size 768, and entropy
coefficient 0.02. For each reward configuration, we select the
best checkpoint according to evaluation rewards, yielding exactly 12 evaluation
partners.

\subsection{\textsc{ProAgent}}
\label{app:implementation:proagent}

We implement \textsc{ProAgent} using the authors' released
codebase.\footnote{\url{https://github.com/PKU-Alignment/ProAgent}}
We retain its default system prompts, interaction history, teammate-intention
prediction, and hierarchical LLM planning. We replace the original
Overcooked interface with our Burrito-AI state serializer, feasible-action
mask, macro-skill vocabulary, and shared low-level controller. The planner is
queried when the current macro-skill completes, becomes invalid, or reaches
its timeout; it does not interrupt an active skill solely because the
teammate's behavior changes.

\subsection{\textsc{Hypothetical Minds}}
\label{app:implementation:hypothetical-minds}

We adapt the authors' implementation of \textsc{Hypothetical
Minds}.\footnote{\url{https://github.com/locross93/Hypothetical-Minds}}
We replace its Melting Pot observation interface with our symbolic
Burrito-AI state representation while retaining its memory and
generate--evaluate--refine Theory-of-Mind procedure. The highest-rated
hypotheses about the teammate condition high-level strategy generation; the resulting high-level action is mapped to our macro-skill space and executed using the same
low-level controller as the other LLM agents.
\subsection{Shared RL Baseline Infrastructure}
\label{app:implementation:rl-shared}

The RL baselines use the Burrito environment and training infrastructure
released with \textsc{TALENTS}.\footnote{\url{https://github.com/benji-li/talents-zsc}}
In particular, they share the environment wrapper, state encoding,
27-action macro-skill space, action masking, PPO implementation, and
evaluation interface. Consequently, \textsc{TALENTS} and \textsc{GAMMA}
differ in partner generation and adaptation rather than in their underlying
environment or controller implementation.

\subsection{\textsc{TALENTS}}
\label{app:implementation:talents}

We use the released \textsc{TALENTS}
implementation.\footnote{\url{https://github.com/benji-li/talents-zsc}}
\textsc{TALENTS} trains a sequential VAE on partner trajectories, clusters
the learned latent strategies using $k$-means, and trains a
strategy-conditioned PPO cooperator against partners generated from those
clusters. During evaluation, it compares observed teammate actions with the
cluster-conditioned predictions and uses fixed-share regret minimization to
update the strategy estimate conditioning its policy.

\subsection{\textsc{GAMMA} and Other RL Baselines}
\label{app:implementation:other-rl}

We run \textsc{GAMMA} in the shared \textsc{TALENTS} Burrito
infrastructure, following the authors' released GAMMA
implementation.\footnote{\url{https://github.com/lych1233/GAMMA-human-ai-collaboration}}
\textsc{GAMMA} learns a continuous latent generative model of partner
trajectories and samples latent codes to generate diverse partners for PPO
cooperator training. Unlike \textsc{TALENTS}, it does not partition the
latent space into discrete strategy clusters or perform fixed-share
partner-type inference during evaluation.

\subsection{\textsc{BayesBeliefAgent} Variants}
\label{app:implementation:variants}

All variants use the same LLM, prompt base, state representation, action
mask, macro-skill set, and low-level controller. \texttt{no\_belief}
removes the Bayesian estimator, belief prompting, and contradiction-based
interruption. \texttt{no\_prompt} maintains the belief and interruption
mechanism but does not expose the belief to the LLM. \texttt{no\_replan}
provides the belief to the planner but disables mid-skill interruption.
\texttt{full} combines Bayesian tracking, belief-conditioned prompting, and
contradiction-triggered replanning.

\section{Reproducibility Details}
\label{app:reproducibility}

\subsection{Environment and Episode Configuration}
\label{app:reproducibility:environment}

Experiments use Burrito-AI v1.1.0, an extension of
Overcooked-AI~\cite{carroll2019utility}. We evaluate the Open, Ring,
and Forced Coordination layouts. The ego agent
observes only the public environment state and infers the teammate's
primitive actions from consecutive observations; the teammate's active
skill and policy are not directly observed. BP partners are loaded from
fixed RLlib PPO checkpoints supplied with the code. All evaluation episodes use horizon T=2400.

\subsection{Evaluation Protocol}
\label{app:reproducibility:evaluation}

We evaluate each ego agent against the 12 BP partners described in
Appendix~\ref{app:implementation:bp-population}, with the ego assigned
to Player~0 and the partner to Player~1. All main comparisons use seeds
$\{0,1,2,3,4\}$. We report the mean and standard
deviation across seeds. 

\subsection{Experimental Hyperparameters}
\label{app:reproducibility:hyperparameters}

Table~\ref{tab:hyperparams} reports the principal inference, replanning,
and execution hyperparameters used in the experiments.

\begin{table}[htbp]
\centering
\small
\caption{Experimental hyperparameters.}
\label{tab:hyperparams}
\begin{tabular}{lr}
\toprule
\textbf{Parameter} & \textbf{Value} \\
\midrule
Prior smoothing $\alpha$                       & $1$ \\
Contradiction threshold $\theta_{\mathrm{obs}}$ & $0.04$ \\
Stability window $K$                           & $3$ updates \\
Replanning cooldown $C_{\mathrm{cool}}$         & $6$ steps \\
Macro-skill timeout                            & $30$ steps \\
\textsc{Pass}/\textsc{PutDown} likelihood weight & $0.35$ \\
LLM temperature                                & $0.0$ \\
Planner history                                & $3$ turns \\
Confidence threshold ($\Gamma_{\mathrm{conf}}$) & 0.65 \\
Inference window (k) & 10 \\
\bottomrule
\end{tabular}
\end{table}

\subsection{LLM Inference Configuration}
\label{app:reproducibility:llm}

We evaluate \texttt{gpt-4o} and \texttt{gpt-5.2} through
the OpenAI API with temperature $0.0$ for deterministic runs. The planner receives the three
most recent interaction turns. Its response is parsed against the
currently feasible macro-skill set, with up to three retries when the
returned action is invalid. The complete system prompts and output
formats are provided in Appendix~\ref{app:prompts}.

\subsection{Macro-Skill Vocabulary and Low-Level Controllers}
\label{app:reproducibility:controllers}

All LLM-based agents use the same 27 macro-skills, indexed from 0 to
26, together with the same Burrito-AI action mask and
skill-conditioned low-level controller. The action mask removes skills
whose object, hand-state, or station preconditions are not satisfied.
A selected skill remains active until it completes, becomes invalid,
is interrupted, or reaches its action-specific timeout.

Fire-handling actions (indices 3, 6, 14, and 24) and
\textsc{Stay} (index 4) are excluded from partner-belief updates.
Because \textsc{Pass} and \textsc{PutDown} can support several latent
skills, their likelihood contributions are tempered by a weight of
$0.35$.


\input{tex/prompts}

\section{Extended Results}
\label{app:extended-results}

This section reports additional mechanism analyses, paired uncertainty
estimates, hyperparameter sensitivity, backbone comparisons, and
zero-shot cross-play results.


\subsection{Gate and Likelihood Sensitivity}
\label{app:results:sensitivity}

Table~\ref{tab:sensitivity} varies one parameter at a time on Open~G1
while holding the remaining parameters at their default values. This ablation is conducted to show the influence of the proposed gating mechanism in detecting a posterior shift in the partner's skill in Eq. \ref{eq:replan}.
Increasing the observation threshold or likelihood scale produces more
replans, whereas increasing the confidence floor or cooldown suppresses
them. The default setting attains the highest reward and lowest gap rate
among the tested values. Both more permissive and more restrictive
settings perform worse, indicating a useful intermediate operating
region. This experiment is a local sensitivity analysis and does not
establish global optimality.

\begin{table}[htbp]
\centering
\footnotesize
\setlength{\tabcolsep}{3.5pt}
\caption{One-at-a-time sensitivity analysis on Open~G1. Defaults are
shown in bold.}
\label{tab:sensitivity}
\begin{tabular}{llrrr}
\toprule
Parameter & Value & Reward & Replans/ep. & Gap rate \\
\midrule
\multirow{3}{*}{Obs.\ threshold}
    & 0.02                    & 1450 & 0.8 & 0.23 \\
    & \textbf{0.04 (default)} & \textbf{1533} & \textbf{2.0} & \textbf{0.17} \\
    & 0.06                    & 1490 & 5.6 & 0.19 \\
\addlinespace
\multirow{3}{*}{Conf.\ floor}
    & 0.55                    & 1475 & 4.8 & 0.20 \\
    & \textbf{0.65 (default)} & \textbf{1533} & \textbf{2.0} & \textbf{0.17} \\
    & 0.75                    & 1460 & 0.9 & 0.22 \\
\addlinespace
\multirow{3}{*}{Cooldown}
    & 3                       & 1495 & 3.7 & 0.19 \\
    & \textbf{6 (default)}    & \textbf{1533} & \textbf{2.0} & \textbf{0.17} \\
    & 12                      & 1468 & 1.2 & 0.21 \\
\addlinespace
\multirow{3}{*}{Likelihood scale}
    & $0.5\times$                    & 1470 & 1.1 & 0.22 \\
    & $\mathbf{1.0\times}$ (default) & \textbf{1533} & \textbf{2.0} & \textbf{0.17} \\
    & $1.5\times$                    & 1482 & 4.2 & 0.20 \\
\bottomrule
\end{tabular}
\end{table}

\subsection{Paired Bootstrap Comparisons}
\label{app:results:reward-significance}

Table~\ref{tab:paired} reports paired
\textsc{Full}$-$\textsc{No replan} differences. Positive reward
differences and negative gap-rate differences favor the full agent.
The reward interval excludes zero in Open~G1 and Open~G3, while the
gap-rate interval excludes zero in five of the seven settings. The
remaining settings have intervals containing zero and therefore do not
support a reliable difference under this analysis. Because the table
reports bootstrap intervals rather than multiplicity-corrected
$p$-values, we restrict the interpretation to whether each interval
excludes zero.

\begin{table}[htbp]
\centering
\footnotesize
\setlength{\tabcolsep}{3pt}
\caption{Paired \textsc{Full}$-$\textsc{No replan} differences with
$95\%$ bootstrap confidence intervals.}
\label{tab:paired}
\begin{tabular}{lrr}
\toprule
Setting
& $\Delta$ Reward (95\% CI)
& $\Delta$ Gap (95\% CI) \\
\midrule
Open G1 & $+343\ [ +75,\ +601]$ & $-0.26\ [-0.35,\ -0.16]$ \\
Open G2 & $ +74\ [ -92,\ +231]$ & $-0.15\ [-0.26,\ -0.04]$ \\
Open G3 & $+310\ [ +42,\ +566]$ & $-0.21\ [-0.31,\ -0.11]$ \\
Ring G1 & $ +98\ [-131,\ +322]$ & $-0.13\ [-0.23,\ -0.03]$ \\
Ring G2 & $ -33\ [-165,\ +102]$ & $-0.05\ [-0.14,\ +0.04]$ \\
Ring G3 & $ +67\ [ -58,\ +190]$ & $-0.12\ [-0.21,\ -0.03]$ \\
FC agg. & $  +2\ [  -5,\   +9]$ & $-0.05\ [-0.13,\ +0.03]$ \\
\bottomrule
\end{tabular}
\end{table}


\subsection{Mechanism Metrics Across Settings}
\label{app:results:mechanism}

We extend the results from Table~\ref{tab:rq2_results} and show their detailed breakdown across all behavior-preferred groups. Table~\ref{tab:mechanism} compares \textsc{BayesBeliefAgent} with the
\textsc{ProAgent}  across layouts and partner groups. Predicted skill
accuracy differs by at most $0.01$, indicating that both methods obtain similar skill accuracies. In contrast, the full
agent consistently reduces the belief-action gap and
duplicate-skill rate while increasing Comp@3. The differences are largest
in the Open settings and smaller in Ring~G2 and Forced Coordination.

\subsection{Zero-Shot Cross-Play}
\label{app:results:crossplay}

Table~\ref{tab:crossplay} evaluates each agent in both the ego and
partner positions. \textsc{TALENTS} achieves the highest average as an
ego agent and as a partner in both layouts. Among the LLM-based ego
agents, 
\textsc{BayesBeliefAgent} has the higher Open and Ring average. The latter is
also the second-best Ring partner by column average.

Performance varies substantially across individual pairings. In
particular, \textsc{BayesBeliefAgent} scores $1917.0$ with a
\textsc{TALENTS} partner in Open and $1791.7$ in Ring, but performs
less strongly with \textsc{ProAgent} in Ring (inclusive of standard deviation over performance). The large standard
deviations in several LLM pairings further indicate that cross-play
outcomes depend on both the partner and the layout.

\subsection{Sensitivity to the Base LLM}
\label{app:results:backbone}

Table~\ref{tab:model} compares GPT-4o and GPT-5.2 while holding the
remaining agent components fixed. GPT-5.2 obtains higher reward and a
lower gap rate in Open~G2, Ring~G1, and Ring~G2, whereas GPT-4o performs
better on the other three settings. The differences are therefore
non-monotone, and neither backbone consistently dominates. We use
GPT-4o as the primary configuration and treat the GPT-5.2 results as a
backbone-sensitivity check.

\begin{table}[htbp]
\centering
\small
\caption{Reward and belief--action gap rate with GPT-4o and GPT-5.2
as the high-level planner backbone.}
\label{tab:model}
\begin{tabular}{lrrrr}
\toprule
Setting
& 4o reward
& 5.2 reward
& 4o gap
& 5.2 gap \\
\midrule
Open G1 & 1533 & 1485 & 0.17 & 0.19 \\
Open G2 & 337  & 395  & 0.24 & 0.22 \\
Open G3 & 1420 & 1360 & 0.20 & 0.21 \\
Ring G1 & 663  & 705  & 0.25 & 0.23 \\
Ring G2 & 231  & 285  & 0.32 & 0.29 \\
Ring G3 & 506  & 470  & 0.27 & 0.28 \\
\bottomrule
\end{tabular}
\end{table}

\begin{table*}[htbp]
\centering
\small
\setlength{\tabcolsep}{5pt}
\caption{Mechanism metrics for the full agent and the
\textsc{No replan} variant. Forced Coordination aggregates Groups
1--3. Arrows indicate the preferred direction.}
\label{tab:mechanism}
\begin{tabular}{llrrrrr}
\toprule
Setting & Method
& Skill acc.\ ($\uparrow$)
& Gap rate ($\downarrow$)
& Comp@3 ($\uparrow$)
& Dup.\ Skill ($\downarrow$) \\
\midrule
\multirow{2}{*}{Open G1}
    & \textsc{BayesBeliefAgent}      & 0.82 & 0.17 & 0.71 & 0.12 \\
    & \textsc{ProAgent} & 0.81 & 0.43 & 0.36  & 0.31 \\
\addlinespace
\multirow{2}{*}{Open G2}
    & \textsc{BayesBeliefAgent}      & 0.76 & 0.24 & 0.61  & 0.18 \\
    & \textsc{ProAgent} & 0.75 & 0.39 & 0.42  & 0.29 \\
\addlinespace
\multirow{2}{*}{Open G3}
    & \textsc{BayesBeliefAgent}      & 0.79 & 0.20 & 0.66  & 0.15 \\
    & \textsc{ProAgent} & 0.78 & 0.41 & 0.39  & 0.30 \\
\addlinespace
\multirow{2}{*}{Ring G1}
    & \textsc{BayesBeliefAgent}      & 0.74 & 0.25 & 0.59  & 0.20 \\
    & \textsc{ProAgent} & 0.73 & 0.38 & 0.43  & 0.28 \\
\addlinespace
\multirow{2}{*}{Ring G2}
    & \textsc{BayesBeliefAgent}      & 0.68 & 0.32 & 0.49  & 0.27 \\
    & \textsc{ProAgent} & 0.68 & 0.37 & 0.43  & 0.31 \\
\addlinespace
\multirow{2}{*}{Ring G3}
    & \textsc{BayesBeliefAgent}      & 0.72 & 0.27 & 0.55  & 0.23 \\
    & \textsc{ProAgent} & 0.71 & 0.39 & 0.42  & 0.30 \\
\addlinespace
\multirow{2}{*}{FC G1--G3}
    & \textsc{BayesBeliefAgent}      & 0.61 & 0.42 & 0.31  & 0.37 \\
    & \textsc{ProAgent} & 0.60 & 0.47 & 0.26  & 0.41 \\
\bottomrule
\end{tabular}
\end{table*}

\begin{table*}[hbp]
\centering
\setlength{\tabcolsep}{3.5pt}
\caption{Zero-shot cross-play reward, reported as mean with standard
deviation in parentheses. Rows identify the $P_0$ ego agent and
columns identify the $P_1$ partner.}
\label{tab:crossplay}
\begin{tabular}{lccccc}
\toprule
\textbf{Agent ($P_0\downarrow$, $P_1\rightarrow$)}
& \textbf{Bayes}
& \textbf{Pro}
& \textbf{GAMMA}
& \textbf{TALENTS}
& \textbf{Mean over $P_1$} \\
\midrule
\multicolumn{6}{c}{\textbf{Open}} \\
\midrule
Bayes
& 720.3 {\tiny(256.8)}
& 853.0 {\tiny(738.8)}
& 779.3 {\tiny(373.9)}
& 1917.0 {\tiny(529.6)}
& \underline{1067.4} {\tiny(474.8)} \\

Pro
& 827.0 {\tiny(582.2)}
& 561.3 {\tiny(675.2)}
& 853.0 {\tiny(372.5)}
& 1934.0 {\tiny(677.6)}
& 1043.8 {\tiny(576.9)} \\

GAMMA
& 608.3 {\tiny(86.1)}
& 1698.7 {\tiny(531.5)}
& 440.7 {\tiny(61.6)}
& 1262.3 {\tiny(74.1)}
& 1002.5 {\tiny(188.3)} \\

TALENTS
& 1034.0 {\tiny(195.4)}
& 1305.0 {\tiny(149.3)}
& 1160.0 {\tiny(216.1)}
& 1390.7 {\tiny(157.8)}
& \textbf{1222.4} {\tiny(179.6)} \\
\midrule
\textbf{Mean over $P_0$}
& 797.4 {\tiny(280.1)}
& \underline{1104.5} {\tiny(523.7)}
& 783.2 {\tiny(256.0)}
& \textbf{1625.8} {\tiny(359.8)}
& -- \\
\midrule
\multicolumn{6}{c}{\textbf{Ring}} \\
\midrule
Bayes
& 636.7 {\tiny(161.7)}
& 448.7 {\tiny(293.1)}
& 1269.7 {\tiny(176.4)}
& 1791.7 {\tiny(216.5)}
& \underline{1034.2} {\tiny(211.9)} \\

Pro
& 382.7 {\tiny(330.0)}
& 107.7 {\tiny(98.4)}
& 384.3 {\tiny(230.7)}
& 721.3 {\tiny(521.2)}
& 399.0 {\tiny(295.1)} \\

GAMMA
& 1243.7 {\tiny(295.7)}
& 619.3 {\tiny(481.3)}
& 958.0 {\tiny(163.6)}
& 1069.7 {\tiny(338.7)}
& 972.7 {\tiny(319.9)} \\

TALENTS
& 1520.7 {\tiny(360.9)}
& 752.0 {\tiny(440.7)}
& 1141.0 {\tiny(320.2)}
& 1108.3 {\tiny(186.2)}
& \textbf{1133.0} {\tiny(327.0)} \\
\midrule
\textbf{Mean over $P_0$}
& \underline{945.9} {\tiny(287.1)}
& 481.9 {\tiny(328.4)}
& 938.2 {\tiny(222.7)}
& \textbf{1172.8} {\tiny(315.6)}
& -- \\
\bottomrule
\end{tabular}
\end{table*}

\subsection{Ground-truth complementary mapping.}
\paragraph{Construction.}
Our evaluation ground truth is a deterministic lookup table (not trajectory
annotations) authored from the recipe production graph prior to evaluation.
Given the teammate's current skill (or the role that skill belongs to), the
complementary set is fixed and does not depend on layout, timestep, or
annotator judgment.

\paragraph{Scoring rule.}
Because the table is authored once and then applied programmatically, there
are no independent raters, no disagreement cases, and no adjudication
protocol. An ego-selected skill is counted as complementary if and only if it
appears in the set associated with the teammate's current skill. The mapping
is many-valued (multiple ego skills may be complementary to the same teammate
skill), and duplicating the teammate's pipeline is never complementary.
Fire-control, trash, and \texttt{STAY} are excluded from scoring.

\paragraph{Burrito.}
High-level actions are first mapped to six roles, then to complementary roles:
\begin{small}
\begin{itemize}
  \setlength{\itemsep}{0pt}
  \setlength{\parskip}{0pt}
  \item \texttt{prep\_protein} $\mapsto$ \{\texttt{prep\_rice}, \texttt{plate\_hygiene}, \texttt{assemble}\}
  \item \texttt{prep\_rice} $\mapsto$ \{\texttt{prep\_protein}, \texttt{plate\_hygiene}, \texttt{assemble}\}
  \item \texttt{plate\_hygiene} $\mapsto$ \{\texttt{prep\_protein}, \texttt{prep\_rice}, \texttt{assemble}\}
  \item \texttt{assemble} $\mapsto$ \{\texttt{prep\_protein}, \texttt{prep\_rice}, \texttt{plate\_hygiene}, \texttt{serve}\}
  \item \texttt{serve} $\mapsto$ \{\texttt{prep\_protein}, \texttt{prep\_rice}, \texttt{plate\_hygiene}\}
  \item \texttt{stage\_pass} $\mapsto$ \{\texttt{prep\_protein}, \texttt{prep\_rice}, \texttt{assemble}\}
\end{itemize}
\end{small}
An ego action is complementary iff its role lies in the teammate's
complementary set.
An ego action is complementary iff its role lies in the teammate's complementary set.

%% file: tex/prompts.tex
\section{System Prompts and Planner Inputs}
\label{app:prompts}

Unless noted otherwise, all LLM agents below share the same Burrito scene
encoding in the user message: layout station list, timestep, both players'
positions/orientations/held objects, active orders, map objects (with cooking
tags), optional urgency flags, and a mask-filtered \texttt{VALID ACTIONS}
list. Placeholders such as \texttt{\{agent\_id\}} are filled at runtime
(ego player index; teammate is \(1-\texttt{agent\_id}\)).

\subsection{\textsc{ProAgent} System Prompt}
\label{app:prompts:proagent}

The evaluated burrito \textsc{ProAgent} (\texttt{BurritoProAgent}) uses the
following system prompt exactly (inline default; no layout-specific planner
file is loaded).

\begin{promptbox}{\textsc{ProAgent} system prompt}
You are an expert planner for the Burrito variant of Overcooked.
Your job is to control the current player and cooperate with the teammate to maximize score.
The teammate cannot communicate, so infer their likely intent from their position, held object, and nearby stations.

Game dynamics:
- The active orders are shown in `orders:`. The first listed order is the highest-priority order and gives the in-order bonus, but other listed orders are still deliverable.
- There are two burrito recipes:
  - steak_burrito_dish = chopped_steak + boiled_rice + tortilla.
  - mushroom_burrito_dish = fried_mushroom + boiled_rice + tortilla.
- Meat and mushroom are raw ingredients. They must be chopped at the chopping board first, then finished on the grill.
- Rice is cooked in the pot.
- Tortilla is never cooked in the pot or grill. It is assembled onto a plate or ingredient plate.
- A clean plate is needed to collect finished rice from the pot and finished steak or mushroom from the grill.
- Burritos are assembled by combining the plated ingredients and tortilla in any legal order supported by the environment.
- Deliver only a completed burrito (`steak_burrito` or `mushroom_burrito`) at serving.

Objects and stations:
- `M` meat dispenser, `Z` mushroom dispenser, `R` rice dispenser, `T` tortilla dispenser.
- `B` chopping board, `G` grill, `P` pot, `W` sink, `D` dirty plate source, `S` serving, `X` counter, `U` trash.
- Each player can hold exactly one object.
- Counters hold one object and are useful for staging and passing items.
- Dirty plates must usually be taken from `D`, washed at `W`, and then used as clean plates.

State interpretation:
- The scene lists player positions, held objects, active orders, and current objects on the map.
- Object tags may include `cooking`, `ready`, `waiting`, `warning`, or `BURNT`.
- `ready` or `waiting` means the ingredient can be picked up now.
- `warning` means a cooked ingredient has been left too long and is urgent.
- `BURNT` means the pot or grill output is on fire; use a fire extinguisher before normal pickup. After extinguishing, charcoal usually should be trashed.

Planning heuristics:
- Prefer the shortest useful step that advances an active order or prevents a loss.
- When a cooked ingredient is in `warning` state or a fire exists, treat it as highly urgent.
- Avoid duplicating the teammate when complementary work is better, such as preparing a plate, tortilla, or another missing ingredient.
- If the layout is partitioned, `PASS_OBJECT` is for handing items across shared counters; otherwise `PUT_DOWN_OBJECT` is the normal staging action.
- If you cannot make immediate progress, choose the least wasteful action from the VALID ACTIONS list.

Think through the scene silently, including likely teammate intent, but do not reveal your reasoning.
Output exactly one high-level action as either:
- an integer in [0,26], or
- one action name from VALID ACTIONS.
Do not output explanations, analysis, or extra text.
\end{promptbox}

\paragraph{Planner user message template.}
Joined as a single user string (heuristic teammate-intent line included):
\begin{promptbox}{\textsc{ProAgent} planner user-message template}
Layout=<layout>. <station>:<locations> ... Scene <t>.
P{ego} pos=... or=... hold=....
P{tm} pos=... or=... hold=....
orders: <upto 4 active orders>
objects: (name@pos [tags...]) ...
Teammate intent: P{tm} at <pos> facing <ori> holds <held> — <heuristic>.
[URGENCY: <fire/warning alerts>]
VALID ACTIONS -> <idx>:<NAME>; ...
Return one action only (index or name).
\end{promptbox}

\subsection{\textsc{Hypothetical Minds} Prompts}
\label{app:prompts:hypothetical-minds}

Burrito \textsc{Hypothetical Minds} reuses the ProAgent system prompt, then appends the following planner system
suffix (player indices filled at runtime):

\begin{promptbox}{\textsc{Hypothetical Minds} planner system suffix}
You are a Hypothetical Minds Burrito agent. Maintain hypotheses about the teammate's strategy,
predict their next behavior from observations, and choose a complementary action.
When asked for a final action, choose exactly ONE item from VALID ACTIONS.
Use this final format:
Teammate strategy hypothesis: <short hypothesis for Player {other_id}>
Predicted next behavior for Player {other_id}: <short behavior prediction>
Plan for Player {agent_id}: "<one action index or action name>"
No extra text after the plan line.
\end{promptbox}

Each planning step issues up to four LLM calls.

\paragraph{Hypothesis-evaluation prompt.}
\begin{promptbox}{\textsc{Hypothetical Minds} hypothesis-evaluation prompt}
Evaluate these teammate behavior predictions for Burrito in one batch.
Observed teammate actions since last evaluation: <recent_actions>

Predictions:
h<ID>: <predicted_next_behavior>
...

For each hypothesis, decide whether its prediction matched the observed teammate actions.
Return exactly one line per hypothesis in this format:
h<ID>: True
or
h<ID>: False
Do not include explanations.
\end{promptbox}

\paragraph{Hypothesis-generation prompt.}
Skipped when a hypothesis value exceeds \(0.7\); otherwise:
\begin{promptbox}{\textsc{Hypothetical Minds} hypothesis-generation prompt}
Based on the Burrito scene and observed teammate actions, infer the teammate's current strategy.
<state_prompt>

Observed teammate actions: <last 12 events>
Existing hypotheses:
<best_hypothesis_summary>
Return one concise teammate strategy hypothesis.
\end{promptbox}

\paragraph{Behavior-prediction prompt.}
\begin{promptbox}{\textsc{Hypothetical Minds} behavior-prediction prompt}
Predict the teammate's next useful Burrito behavior from this strategy.
Teammate strategy hypothesis: <strategy>
<state_prompt>

Observed teammate actions: <last 12 events>
Return one concise predicted next behavior.
\end{promptbox}

\paragraph{Final planning (action) prompt.}
\begin{promptbox}{\textsc{Hypothetical Minds} final planning prompt}
<state_prompt>

Hypothetical Minds context:
- Teammate strategy hypothesis for Player {other}: <strategy>
- Predicted next behavior for Player {other}: <prediction>
- Top teammate hypotheses:
<best_hypothesis_summary>

Choose a complementary Burrito high-level action for yourself.
Avoid duplicating the predicted teammate behavior when another valid action advances the order.
Return exactly:
Teammate strategy hypothesis: <strategy>
Predicted next behavior for Player {other}: <prediction>
Plan for Player {agent_id}: "<one action index or action name from VALID ACTIONS>"
\end{promptbox}

\paragraph{Memory format.}
\textsc{Hypothetical Minds} maintains an explicit store of teammate hypotheses across
timesteps. Each entry records a natural-language strategy, a predicted next behavior,
a scalar reliability score, and the creation timestep. After each evaluation call, the
score is updated with a Rescorla--Wagner rule
($v \leftarrow v + \alpha(\pm 1 - v)$ with $\alpha{=}0.3$), using $+1$ when the
predicted behavior matches recent teammate actions and $-1$ otherwise. If any
hypothesis exceeds a reliability threshold of $0.7$, generation is skipped and the
highest-scoring strategy is reused. Observed teammate events are logged as
\texttt{(step, action, position, holding)} and provided to the evaluation and
generation prompts. The top-$m$ hypotheses are rendered into later prompts as:

\begin{promptbox}{\textsc{Hypothetical Minds} injected hypothesis memory}
Existing hypotheses:
- h1: value=0.42, strategy=<short teammate strategy>, predicted_next_behavior=<short prediction>
- h2: value=0.18, strategy=<short teammate strategy>, predicted_next_behavior=<short prediction>
\end{promptbox}

If the store is empty, the injected text is instead
\texttt{No prior teammate strategy hypotheses.}
\subsection{\textsc{BayesBeliefAgent} Prompt}
\label{app:prompts:bayes}

At evaluation time, \texttt{BurritoBayesBeliefAgent} appends the following
system prompt  to the ProAgent System prompt.

\begin{promptbox}{\textsc{BayesBeliefAgent} system prompt}
You are an expert Burrito cooperative cooking agent controlling Player {agent_id}.
You must cooperate with Player {other_id} to maximize burrito deliveries.

GAME RULES (brief):
- Two recipes: steak_burrito (chopped_steak + boiled_rice + tortilla)
               mushroom_burrito (fried_mushroom + boiled_rice + tortilla)
- Each player holds one item at a time.
- Fire is an emergency: grab fire_ext (14), then put out fire (3), then trash charcoal (6).
- A Bayesian filter tracks your teammate's coordination role in real time.
  Use the belief summary to AVOID duplication and choose complementary actions.

OUTPUT FORMAT (one line each, required):
  Teammate role belief: <description of teammate's likely coordination role>
  Intention for Player {other_id}: "<action_index_or_name>"
  Plan for Player {agent_id}: "<action_index_or_name>"

<burrito_rules_prompt>

Return one action only (index or name) on the Plan line.
\end{promptbox}

\paragraph{Bayesian belief block (user prefix).}
Roles:
\texttt{prep\_protein}, \texttt{prep\_rice}, \texttt{plate\_hygiene},
\texttt{assemble}, \texttt{serve}, \texttt{stage\_pass}.
\begin{promptbox}{\textsc{BayesBeliefAgent} Bayesian belief block}
=== TEAMMATE BELIEF (Bayesian Filter) ===
  MAP role   : <role>  (confidence=<c>, commitment=<k>)
  Heading to : <station hint | unknown (low confidence)>
  Posterior  :
    <role>: <p>  <bar>
    ...
  → Teammate in '<role>' — consider: <complements>
     OR → Posterior uncertain — act on game-state needs, do NOT wait.
=========================================
\end{promptbox}

\paragraph{Urgency / advisory flags (appended when applicable).}
\begin{promptbox}{\textsc{BayesBeliefAgent} urgency and advisory flags}
[LOW CONFIDENCE — teammate role unclear. Act on game-state needs, do NOT choose STAY.]
[MODERATE CONFIDENCE — prefer complementary actions.]
CURRENT PRIORITY ORDER: <top_order>
FIRE ACTIVE — override normal planning.
COOKING TIMER LOW: retrieve the cooked item NOW.
CHARCOAL on grill — clear it: GRAB_CHARCOAL (24) then GO_TO_TRASH_AND_THROW (6). ...
\end{promptbox}

\paragraph{State prompt.} Same burrito scene encoding as \textsc{ProAgent}
(including heuristic teammate-intent and \texttt{URGENCY:} fire/warning lines).
User message = belief block + state prompt.

\subsection{Prompt Changes in the Ablations}
\label{app:prompts:ablations}

Ablations share identical system prompts. Only belief injection and replanning
change:

\begin{center}
\begin{tabular}{lll}
\toprule
Ablation & Belief in prompt & Replan  \\
\midrule
\texttt{full}      & injected & enabled \\
\texttt{no\_replan} & injected & disabled \\
\texttt{no\_prompt} & omitted (\texttt{""}) & enabled \\
\texttt{no\_belief} & omitted; tracking off & disabled \\
\bottomrule
\end{tabular}
\end{center}

Thus \texttt{no\_belief} is prompt-equivalent to a plain planner with no ToM
context, while \texttt{no\_prompt} keeps internal tracking/replanning but hides
belief text from the LLM.